\documentclass[amssymb,floatfix,twocolumn,a4paper]{revtex4} 
\usepackage[dvips]{graphicx} 
\usepackage[latin1]{inputenc}
\usepackage{color,rotating}
\usepackage{amsmath} 
\usepackage{amssymb} 
\usepackage[english]{babel}
\usepackage{dwellTimeNotation}
\begin{document}
\title{
Dwell time symmetry in random walks and molecular motors} 

\pagestyle{myheadings}
\markright{Dwell time symmetry}
\author{Martin Lind\'en}
\email{linden@kth.se}
\author{Mats Wallin}
\email{wallin@kth.se}
\affiliation{Department of Theoretical Physics, Royal Institute of
Technology (KTH), AlbaNova, 10691 Stockholm, Sweden}

\begin{abstract}
The statistics of steps and dwell times in reversible molecular motors
differ from those of cycle completion in enzyme kinetics.  The reason
is that a step is only one of several transitions in the
mechanochemical cycle. As a result, theoretical results for cycle
completion in enzyme kinetics do not apply to stepping data.  To allow
correct parameter estimation, and to guide data analysis and
experiment design, a theoretical treatment is needed that takes this
observation into account.  In this paper, we model the distribution of
dwell times and number of forward and backward steps using first
passage processes, based on the assumption that forward and backward
steps correspond to different directions of the same transition.  We
extend recent results for systems with a single cycle and consider the
full dwell time distributions as well as models with multiple
pathways, detectable substeps, and detachments.  Our main results are
a symmetry relation for the dwell time distributions in reversible
motors, and a relation between certain relative step frequencies and
the free energy per cycle.  We demonstrate our results by analyzing
recent stepping data for a bacterial flagellar motor, and discuss the
implications for the efficiency and reversibility of the
force-generating subunits.\\
\emph{Key words:} motor proteins; single molecule kinetics; enzyme
kinetics; flagellar motor; Markov process; non-equilibrium
fluctuations.

\end{abstract}
\maketitle 

\section*{Introduction}
Progress in single molecule techniques has enabled observations of
single steps in many motor proteins, and accurate measurement of the
distribution of dwell times, i.e., the periods of little or no motion
between steps. Examples are the forward and backward steps of
processive molecular motors like kinesin
\citep{carter05,nishiyama02,taniguchi05,guydosh06}, myosin V
\citep{uemura04,rief00,clemen05,gebhardt06}, cytoplasmic dynein
\citep{toba06} or RNA polymerase \citep{abbondanzieri05}, and stepwise
rotations in ATP synthase
\citep{diez04,ueno05,yasuda98,yasuda01,shimabukuro03,shimabukuro05,nishizaka04}
and the flagellar motor \citep{sowa05}.  Observations of steps and
dwell time distributions offer a route to gain insight into the
microscopic mechanism and detailed motion of such systems, beyond what
is available through knowledge of average turnover rates alone.

Dwell times are examples of first passage times, which have been
extensively studied in the theory of random walks
\citep{vankampen}. An important modeling step is therefore to
formulate a first passage problem that describes the experimental
situation.  Close examination of step trajectories with high time
resolution from several motor proteins
\citep{guydosh06,taniguchi05,nishiyama02,carter05,uemura04,rief00,clemen05,
gebhardt06,diez04,ueno05,yasuda98,yasuda01,shimabukuro03,shimabukuro05,
nishizaka04,sowa05,toba06} reveals that the steps are very rapid
events compared to typical dwell times. In a discrete-state
description, it is therefore reasonable to identify a step with a
single transition, and the direction of the step with the direction of
that transition \citep{tsygankov06a,qian97}.  The identification of
steps with single transitions follows naturally from the assumption
that each state has a well-defined average position, and is a basic
assumption in this paper.

Motor proteins are cyclic enzymes, but contrary to earlier assumptions
\citep{kolomeisky03,kolomeisky05}, the steps and waiting times
obtained in many single-molecule experiments are not well described in
terms of the cycle completions in enzyme kinetics.  This was recently
demonstrated for models where a single cycle accounts for both forward
and backward steps \citep{tsygankov06a}.  In this case, the average
number of forward and backward steps differ from the average number of
completed forward and backward cycles, and the average dwell times
between steps differ from the average cycle completion times. The
differences can be significant even in conditions where almost no
backward steps occur.  The basic reason for this difference is that a
motor with one or more intermediate states per cycle is in a different
state immediately after forward and backward steps. Therefore, an
experimental trajectory of forward and backward steps does not give
explicit information about completed forward and backward cycles.

The different states obtained just after forward and backward steps
also implies that consecutive step directions are correlated
\citep{tsygankov06a}. This prediction will be confirmed in a later
section, where we analyze stepping data from \citet{sowa05} for a
flagellar motor, and find the clear step-step correlations shown in
\Fig{flagellaPC}(b). In contrast, consecutive cycle completions are
statistically independent \citep{qian06,wang06}.

The observation that steps are correlated has important implications
for the interpretation of stepping experiments, and also motivates
further theoretical study of stepping statistics that goes beyond the
assumption \citep{shaevitz05,santos05,svoboda94} of independent steps
and dwell times.  In this paper, we extend the theory of
\citet{tsygankov06a} in two ways. First, we consider the distributions
of conditional dwell times instead of mean values. Second, we consider
a larger class of models, including motor detachments, substeps, and
multiple pathways.

\begin{figure}[b]\begin{center}
\includegraphics{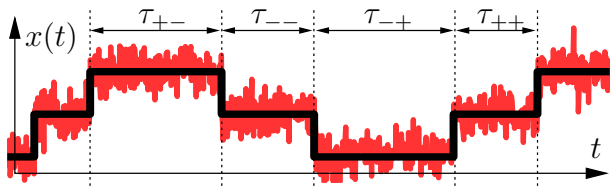}
\end{center}
  \caption{Example of the four different conditional dwell times in a
    model with one step per enzymatic cycle (synthetic data). The
    coordinate $x(t)$ could be position of a linear motor, net
    rotation of a rotary motor, or the net number of a substrate or
    product in an enzymatic reaction, and changes stepwise in steps of
    size $d$. The staircase line is an idealized running average.
    }\label{dwelltypes}
\end{figure}

Our main results are a distribution symmetry for conditional dwell
times, and a simple relation between the dissipated free energy per
cycle, $\Delta G$, and certain conditional stepping probabilities. For
a large number of models relevant to describe reversible motor
proteins, we get
\begin{subequations}\label{mainresult}
  \begin{align}
      \rff(t)=\rbb(t)&\Leftrightarrow P_{++}(t)=P_{--}(t),\label{mainresult1}\\
      \piff/\pibb&=e^{-\Delta G/\kBT}.\label{mainresult2}
  \end{align}
\end{subequations}
Here, $\rff(t)$ and $\rbb(t)$ are the probability density functions
for the conditional dwell times between two consecutive forward and
backward steps respectively,
$P_{\pm\pm}(t)=\int_0^t\Cpdf{\pm\pm}(t)dt$ are the corresponding
integrated probability functions, $\piff$ is the probability that a
forward step is followed by another forward step, and $\pibb$ that a
backward step is followed by another backward step.  The different
types of dwell times are illustrated in \Fig{dwelltypes}, and
\Eq{mainresult1} simply means that the conditional dwell times $\tff$
and $\tbb$ are random variables with equal distributions. However,
\Eq{mainresult} does not say anything about the distributions of dwell
times between steps of different directions, which in general have
different distributions.
Equation \eqref{mainresult} holds also at finite average velocity. For
example, a forward-moving motor will take mostly forward steps, but
might eventually produce two consecutive backward steps. The dwell
times between such $++$ and $--$ step pairs have equal distributions,
although the probability of observing two consecutive backward steps
might be very small for a motor with strong forward bias.

In practice, the need to observe a significant number of $++$ and $--$
events to test or apply \Eq{mainresult} can be an experimental
challenge. For example, ATP driven motors like kinesin or myosin V
typically have $\Delta G_\text{ATP}\approx -25 \kBT$
\emph{in vivo} \cite{howard}. This means that $--$ events are very
rare (see \Eq{BBestimate}), unless an external load is applied
\cite{carter05,nishiyama02,taniguchi05,guydosh06,uemura04,rief00,clemen05,gebhardt06}.

To extract qualitative information out of an experimental test of
dwell time symmetry, one needs a characterization of the class of
models which satisfy \Eq{mainresult}. We try to formulate a general
characterization of this class.  This means that some of the models
within this characterization will not be realistic descriptions of
biological systems. On the other hand, a large number of candidate
models can be excluded if dwell time symmetry is not observed in some
system.  We find two sufficient assumptions for \Eq{mainresult} to
hold, which we call \emph{strong coupling} and the \emph{bottleneck
property}. Within the discrete state modeling framework that we use,
these assumptions are easy to formulate, but can also be disposed of
\cite{bustamante01}.

The bottleneck property is an assumption about the model topology. It
means that both forward and backward steps correspond to transitions
to or from a single state in the mechanochemical cycle, which we call
the bottleneck state. This ensures that the state of the system is
uniquely determined after each observed step, independent of previous
step directions.  The bottle-neck property alone results in a
particularly simple form for the step-step correlations, given in
\Eq{kc}, which can be tested experimentally.

The assumption of strong coupling, defined mathematically in
\Eq{micrev}, is related to microscopic reversibility. Physically, it
means that there is no futile free energy dissipation in the
mechanochemical cycle.  In addition to tight coupling in the usual
sense \citep{bustamante01}, i.e., a one-to-one correspondence between
fuel consumption and forward steps, it includes the assumption that
backward steps is tightly coupled to synthesis of fuel. As a result,
motors with strong coupling are effectively one-dimensional systems,
in the sense that only one reaction coordinate, e.g., position, is
needed to describe their operation.

Strong coupling might not necessarily hold for all motors.  For
example, the forward steps of Myosin V are tightly coupled to ATP
hydrolysis, while backward steps independent of ATP binding was
recently reported \citep{gebhardt06}. Another example is kinesin, for
which ATP binding during steps in both directions has been reported
\citep{nishiyama02,carter05}. On the other hand, strong coupling might
be possible in some rotary motors, at least under certain conditions
\citep{itoh04,nishizaka04,berg00,bustamante01}.

Several recent works \citep{milescu06,milescu06b,mckinney06,smith01}
use Hidden Markov models to estimate kinetic parameters directly from
experimental trajectories, and thus in principle utilize all
information in the data and bypass the difficulties associated with
step detection. However, this approach does not replace the need for a
theoretical understanding of the capabilities and limitations of the
underlying stochastic models. Our results should provide useful
guidance when applying these techniques.

Equation \eqref{mainresult} corresponds to a similar result in cycle
kinetics \citep{qian06,wang06,kolomeisky05,hill89}. Using the symbol
``$\tilde{\phantom{a}}$'' to denote cycle time properties, one has
\begin{equation}\label{cycleresult}
  \tilde \rho_+(t)=\tilde \rho_-(t),\quad
  \tilde\pi_+/\tilde\pi_-=\tilde{J}_+/\tilde{J}_-= e^{-\Delta G/\kBT},
\end{equation}
where $\tilde \rho_\pm(t)$ are the probability density functions for
waiting times before forward ($+$) or backward ($-$) cycle
completions, $\tilde\pi_\pm$ are the relative frequencies of completed
forward and backward cycles respectively, and $\tilde{J}_\pm$ are
one-way cycle fluxes \citep{hill89}.  Equation \eqref{cycleresult} is
based on similar assumptions as \Eq{mainresult}
\cite{qian06,wang06}. However, the waiting times in \Eq{cycleresult}
do not describe the experimental dwell times observed in stepping
experiments on motor proteins \citep{tsygankov06a}. Distinguishing
between cycle completion times and dwell times results in significant
differences when modeling and interpreting stepping data.

One example of such a difference is the possibility to calculate
$\Delta G$ from observations of forward and backward steps. In this
case, using the cycle completion result in \Eq{cycleresult} on
stepping data can give large systematic errors in the estimated
$\Delta G$, which can be avoided if \Eq{mainresult} is used instead.

A related example concerns how to interpret the ratio of observed
forward and backward steps, which has been measured for kinesin over a
range of forces and ATP concentrations
\citep{nishiyama02,carter05,taniguchi05}. The force dependence of this
ratio can be described as roughly proportional
\mbox{$\exp(-cF_xd/\kBT)$}, where $d=8.2$ nm is the step length, $F_x$
is the applied load, and $c$ is a numerical factor significantly
smaller than unity.  The dissipated free energy per cycle depends on
applied load as
\begin{equation}\label{dGFx}
  \Delta G(F_x)=F_{\!\!x}d+\Delta G(0).
\end{equation}
This means that \Eq{cycleresult} predicts $c=1$ if steps are assumed
to correspond to completed cycles. In contrast, \Eq{mainresult} gives
no general reason to even expect an exponential behavior of the ratio
of forward and backward steps \citep{tsygankov06a}, and suggests that
$\pi_{++}/\pi_{--}$ is a more relevant quantity to study.  As
mentioned above, kinesin might not satisfy strong coupling, in which
case one must look at more complicated models. In any case, making the
distinction between steps and completed cycles is clearly important in
order to interpret the experiments correctly.

The rest of the paper is organized as follows. In the next section,
we sketch a derivation of our results for the models studied in
Refs.~\citep{tsygankov06a,qian97}.  We also discuss step-step
correlations.  After that, we generalize our results to a broader
class of models, and discuss detachments and multiple pathways in the
mechanochemical cycle. We also generalize \Eq{mainresult} to the case
of several detectable substeps.  To illustrate our results, we then
analyze stepping data for the flagellar motor, before the concluding
discussion. Detailed derivations are given in the appendices.
\section*{Sequential Models}\label{sequentialModels}
\begin{figure}[b]
\begin{center}
\includegraphics{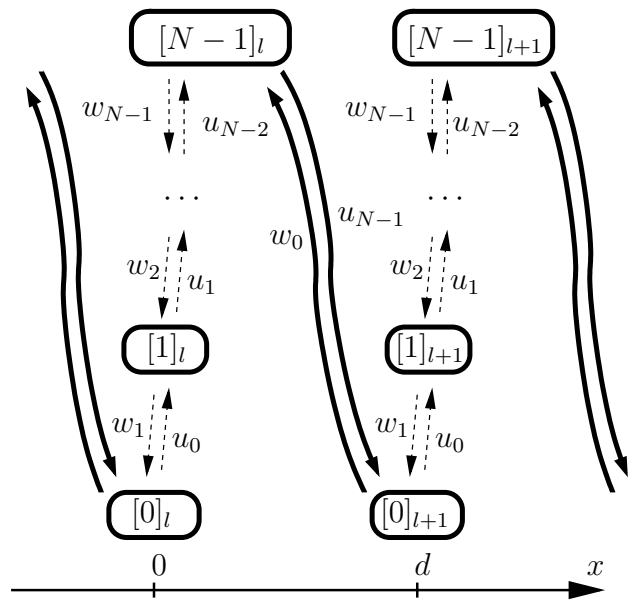}
\end{center}
  \caption{A simple sequential model of an enzyme or motor protein
    with $N$ states per cycle, $[0]_l,[2]_l,\ldots,[N-1]_l$. The model
    can be viewed as a biased random walk on the periodic lattice of
    states.  Numbers in brackets represent different states, and
    indices different cycles. State $[k]_l$ denotes the same
    structural state as $[k+N]_l\equiv [k]_{l+1}$, but with position
    or angular orientation $x$ (for linear and rotary motors
    respectively) shifted by the step length $d$. Arrows indicate the
    allowed transitions between neighbor states with transition rates
    $u_j$, $w_j$.  Fat solid arrows denote the major conformational
    changes that produce observable steps like those in
    \Fig{dwelltypes}. The other transitions are indicated by dashed
    arrows. In a motor protein, these transitions could for example be
    binding or release of ligands, or small conformational changes
    that are hidden in the experimental noise.  }\label{seqPath}
\end{figure}

In this section, we derive \Eq{mainresult} and some related results in
the simplest case, i.e., a sequential model.

Consider a model with a single cycle consisting of a sequence of $N$
states (see \Fig{seqPath}), where we defined $[k]_l$ to denote state
$k$ in cycle $l$, associated with position $ld$.  This is a basic
model in enzyme kinetics \citep{hill89,qian06} and has been used to
describe motor proteins like kinesin
\citep{fisher01,kolomeisky05,fisher05}, \mbox{myosin V}
\citep{kolomeisky03} and F$_1$-ATPase \citep{gao05}. By construction,
it satisfies both strong coupling and the bottleneck property.

We denote the forward (backward) transition rates from state $j$ to
adjacent states by $u_j$ ($w_j$) as indicated in \Fig{seqPath}. The
transition rates are positive and periodic, i.e., $u_{j+N}=u_j$,
$w_{j+N}=w_j$, and possibly functions of external loads and
concentrations of various species in the surrounding solution. For the
purpose of this discussion, the rates are assumed to be arbitrary
positive constants, some of which can be tuned experimentally.

We assume that transitions between different cycles,
$[N-1]_l\rightleftharpoons[0]_{l+1}$, produces observable forward
($+$) or backward ($-$) steps.  If the model describes an enzyme, the
observable step could be the release or uptake of a product or
substrate molecule.

Counting steps is different from counting cycle completions, which can
be illustrated with the following thought experiment. Consider the
sequential model in \Fig{seqPath} with $N=2$ states per cycle, and a
trajectory where the motor goes through the states
$[0]_0\to[1]_{-1}\to[0]_0\to[1]_0\to[0]_1$. This completes one forward
cycle from $[0]_0$ to $[0]_1$, but produces three steps: one backward
($[0]\to[1]_{-1}$), followed by two forward ($[0]_{-1}\to[0]_0$, and
$[1]_0\to[0]_1$). Due to such events, neither the number of steps nor
the dwell times between steps are accurately described in terms of
cycle completions. A different treatment is needed.

The time evolution of the system is a random walk on the periodic
one-dimensional lattice of states, where the average velocity may be
nonzero. To simplify the notation, we now use $j$ to denote a state in
any cycle, with the convention that $j$ and $j+N$ are equivalent
states in different positions.  The probability $q_j(t)$ to be in
state $j$ at time $t$ evolves according to a Master equation
\citep{vankampen}, in this case
\begin{equation}
  \partial_tq_j(t)=u_{j-1}q_{j-1}(t)+w_{j+1}q_{j+1}(t)
  -(u_j+w_j)q_j(t).
\end{equation}
State $j$ has free energy $G_j$, and according to detailed balance,
the free energy difference between two adjacent states are related to
the transition rates through
$u_j/w_{j+1}=e^{-(G_{j+1}-G_j)/\kBT}$. From the periodicity of the
rates we find
\begin{equation}
G_{j+N}=G_j+\Delta G,\quad e^{-\Delta
G/\kBT}=\prod_{i=0}^{N-1}\frac{u_{i}}{w_{i+1}}.
\end{equation}
The direction of the average drift is positive (to the right in Fig.\
\ref{seqPath}) if the dissipated free energy per cycle, $\Delta G$, is
negative, and zero if $\Delta G=0$ \citep{fisher99b,hill89}.

Tracking the position produces a series of forward ($+$) and backward
($-$) steps separated by random dwell times, as sketched in
\Fig{dwelltypes}.  Our aim is to describe the statistics of such a
trajectory, i.e., the fluctuations of the dwell times and the number
of forward and backward steps, which can then be compared with
experimental data.

Following Ref.~\citep{tsygankov06a}, we introduce the pairwise
splitting probabilities $\piff$, $\pifb$, $\pibf$, and $\pibb$, where
$\piff$ and $\pifb$ is the probability that a forward step is followed
by a forward step or a backward step, respectively, and similar for
$\piff,\pibb$. The splitting probabilities satisfy
\begin{equation}\label{splitNormalization}
  \piff+\pifb=\pibf+\pibb=1.
\end{equation}
Similarly, we introduce random variables $\tff$, $\tfb$, $\tbf$, and
$\tbb$ for the conditional dwell times, where $\tff$ is the dwell time
between two consecutive forward steps, and so on, as illustrated in
\Fig{dwelltypes}. Once the proper first passage problem to describe
the dwell times has been formulated, explicit expressions for the
splitting probabilities and mean conditional dwell times can be
computed for arbitrary $N$ \citep{tsygankov06a}, using standard
methods \citep{vankampen}.
\subsection*{The First Passage Problem}
We now describe the first passage problem for dwell times in
sequential models, which we then use to derive \Eq{mainresult}.  The
dwell time symmetry is due to a one-to-one mapping between each $--$
event, i.e., two consecutive backward steps and the dwell time between
them, and a corresponding $++$ event. This is the key observation for
the generalization of \Eq{mainresult} to more complex models in later
sections.

We model the splitting probabilities and probability distribution
functions for the dwell times as a first escape problem from the
interval of states $0,1,\ldots,N-1$. One way to approach this problem
is the approach with absorbing boundaries \citep{vankampen}, which
means solving a reduced Master equation for the states
$0,1,\ldots,N-1$, with absorbing boundaries at both ends
\citep{tsygankov06a,qian97}.  From the assumption that steps are
produced by the transition $[N-1]_l\rightleftharpoons[0]_{l+1}$, it
follows that immediately after a $+$ step, the system is in the state
$0$ ($\pm lN$), and just after a $-$ step, it is in the state $N-1$
($\pm lN$).  

We write the reduced Master equation in matrix form,
\mbox{$\partial_t\vektor{q}=\matris{M}\vektor{q}$}, where the matrix
$\matris{M}$ has elements
\begin{equation}\label{seqMelements}
    M_{ij}=u_j\delta_{i,j+1}+w_j\delta_{i,j-1}-(u_j+w_j)\delta_{ij},
\end{equation}
and $0\le i,j\le N-1$. The element $\matris{M}_{ij}$ is the transition
rate from state $j$ to state $i$. The probability functions are given
by the outgoing probability current,
\begin{align}\label{seqPFdef}
    \pi_{\pm+}P_{\pm+}(t)&=u_{N-1}\int_0^tq_{N-1}(t)\du t\\
    \pi_{\pm-}P_{\pm-}(t)&=w_{0}\int_0^tq_{0}(t)\du t,
\end{align}
where $\pm$ indicates an initial condition $q_j(0)$, given by the
direction the of the previous step (see \Eq{seqIC}). Normalization of
the distributions requires that
\begin{equation}\label{Pnorm}
  \lim_{t\to\infty}P_{\pm\pm}(t)=\int_0^\infty\rho_{\pm\pm}(t)\,dt=1.
\end{equation}
\subsection*{Dwell Time Symmetry}
To derive \Eq{mainresult}, we show that the Taylor series of $\piff
P_{++}(t)$ and $\pibb P_{--}(t)$ are identical up to a factor
$e^{\Delta G/\kBT}$, from which \Eq{mainresult} follows.  The actual
calculation is given in appendix A, and we now discuss some of its
consequences.

First, it is interesting to note that the periodicity of the model is
not necessary for the dwell time symmetry, only to get the simple
relation between $\piff$, $\pibb$ and $\Delta G$ in \Eq{mainresult2}.

Second, the $++/--$ probability distribution has a simple closed form
for (periodic) sequential models. As we show in appendix B,
\begin{equation}\label{seqdist}
  \rff(t)=\rbb(t)=
  (-1)^N\prod_{j=1}^{N}\lambda_j\sum_{k=1}^{N}e^{\lambda_kt}
  \prod_{m\ne k}(\lambda_k-\lambda_m)^{-1},
\end{equation}
where the $\lambda_j$ are the eigenvalues of the matrix $\matris{M}$.
This is the distribution of a sum $N$ of independent exponential
random variables with mean values $\lambda_1^{-1},
\lambda_2^{-1},\ldots, \lambda_N^{-1}$.  

Explicit expressions for all four conditional dwell time distributions
for the case $N=3$ are given in Ref.~\cite{qian97}. As expected from
the calculations in appendix B, there is no simple relation between
$\rfb(t)$ and $\rbf(t)$.

Equation \eqref{seqdist} is only valid if the eigenvalues $\lambda_j$
are distinct. This is the generic situation, since there are no
symmetries or other reasons to expect degeneracies. However,
\Eq{mainresult} is valid also in the degenerate case.

\begin{figure}[b]
\begin{center}
\includegraphics{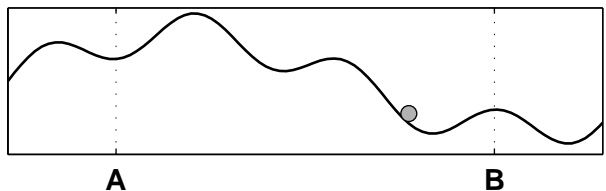}
\end{center}
  \caption{The equations of motion for a particle (gray circle)
    diffusing in an arbitrary potential (solid curve) can be
    discretized to a sequential model \citep{wang03}.  In the
    continuum limit, the dwell time symmetry for the discretization
    translates to equal distributions for the two last-touch
    first-touch waiting times between the (arbitrary) positions A and
    B. }\label{contLim}
\end{figure}

Finally, it is worth noting that the sequential models in this section
can also be obtained as a discretization of overdamped one-dimensional
diffusion \citep{wang03} in an arbitrary potential $U(x)$ between two
points A and B (see \Fig{contLim}). In this case, our results mean
that the waiting time between last touch at A and first touch at B,
and the reverse waiting time between last touch at B and first touch
at A, have equal distributions. This pair of first passage problems
was previously studied by \citet{bier99}.

\subsection*{Step Directions as a Markov Chain}\label{FBsteps}
The fact that the motor is in a different state immediately after
steps in different directions means that the step directions might be
correlated. In this section, we formulate this observation
mathematically, and point out some experimentally relevant
consequences.

Let $\pif(k)$ and $\pib(k)=1-\pif(k)$ be the probabilities that step
$k$ in a trajectory is a forward or backward step respectively. In the
simplest case of no substeps or detachments, the definitions of the
pairwise splitting probabilities give
\begin{equation}\label{stepMarkov}
  \left(\begin{array}{c}\pif(k+1)\\\pib(k+1)\end{array}\right)=
  \left(\begin{array}{cc}\piff&\pibf\\\pifb&\pibb\end{array}\right)
  \left(\begin{array}{c}\pif(k)\\\pib(k)\end{array}\right).
\end{equation}
Equation \eqref{stepMarkov} describes the sequence of forward and
backward steps as a Markov chain \citep{vankampen,tsygankov06a}, and
the normalization constraints in \Eq{splitNormalization} leave two
independent parameters in the $2\times 2$ transition matrix.

The average frequencies $\pi_\pm^*$ of forward and backward steps have
been measured for several motor proteins
\citep{nishiyama02,carter05,uemura04,toba06}. In our model, those
frequencies are given by the stationary distribution of
\Eq{stepMarkov} \citep{tsygankov06a},
\begin{equation}\label{stationarySteps}
  \begin{split}
  \pif^*&=\frac{\pibf}{\pifb+\pibf}=\frac{1-\pibb}{2-\piff-\pibb},\\
  \pib^*&=\frac{\pifb}{\pifb+\pibf}=\frac{1-\piff}{2-\piff-\pibb}.
  \end{split}
\end{equation}
In contrast to the ratio $\tilde{\pi}_+/\tilde{\pi}_-$ of the number
of forward and backward cycles in \Eq{cycleresult}, the ratio
$\pif^*/\pib^*$ of forward and backward steps is in general not equal
to $\exp(-\Delta G/\kBT)$. Looking back at \Eq{dGFx}, we see no
general reason to expect the ratio of forward and backward steps to
depend exponentially on the applied load.

Several experiments have divided the dwell times according to the
direction of the following step
\citep{nishiyama02,carter05,taniguchi05}.  Our model gives the forward
and backward dwell time probability density functions as
\begin{equation}\begin{split}
    \Cpdf{+}^*(t)&=\piff\rff(t)+\pibf\rfb(t),\\
    \Cpdf{-}^*(t)&=\pifb\rbf(t)+\pibb\rbb(t).
  \end{split}\end{equation}
Since $\rfb(t)$ and $\rbf(t)$ are in general not related in a
simple way, one should not expect equal dwell times before forward and
backward steps, in contrast to the case of forward and backward cycle
completion times \citep{kolomeisky05,qian06,wang06}.

Equation \eqref{stepMarkov} implies that step directions are
correlated, and gives the step-step correlation function
\begin{equation}\label{kc}
  C(n)=\mean{s_is_{i+n}}-\mean{s_i}\mean{s_{i+n}}=C(0)\gamma^{|n|},
\end{equation}
where $s_m=\pm 1$ indicates the direction of step $m$, and
\begin{equation}\label{gamma}
  \gamma=\piff+\pibb-1=1-\pifb-\pibf
\end{equation}
is an eigenvalue of the transition matrix in \Eq{stepMarkov}
\citep{tsygankov06a}. The other eigenvalue is one.

Note that $|\gamma|<1$ since $\pi_{\pm\pm}<1$.  In addition, it is
reasonable to expect $\piff\le\pibf$, since a forward step following a
backward step can be accomplished by a single transition, while the
whole cycle must be passed before a forward step is followed by
another forward step.  Inserting $\piff\le\pibf$ in \Eq{gamma}, and
using the normalization in \Eq{splitNormalization}, we get
\begin{equation}\label{gammaLimit}
  \gamma=\piff+\pibb-1\le\pibf+\pibb-1=0.
\end{equation}
Apparently, we should expect negatively correlated steps,
$-1<\gamma\le 0$.  Uncorrelated steps ($\gamma=0$) occur in systems
with $N=1$, i.e., simple random walks, as well as some special cases
of the extended models in the next section.

An upper bound for the number $n_{--}$ of $--$ events is useful to
estimate the regime where our results can be applied in practice.  A
rough estimate is given by
\begin{equation}\label{BBestimate}
  n_{--}/n_\text{tot.} = \pi^*_-\pibb =\pi^*_-\piff e^{\Delta G/\kBT}
  < e^{\Delta G/\kBT}.
\end{equation}
Hence, the number of steps $n_\text{tot}$ required to get adequate
statistics grows at least exponentially with decreasing $\Delta G$.

\section*{Extended Models}\label{extensions}
Beyond the simplest descriptions in terms of a sequential model for
the dominating pathway (if there is one), more complicated situations
with multiple pathways are possible
\citep{kolomeisky01,uemura04,baker04,vilfan05,lan05,xing05}. A general
(coarse-grained) description of a motor protein would keep track of
both the position and the consumption of fuel molecules, which
requires the model to include an effective chemical coordinate in
addition to the position of the motor \citep{bustamante01}. Are the
dwell time symmetry properties of sequential models valid for more
general cases as well?  As we will see next, the answer is yes for a
large class of models, in which the motion is still effectively
one-dimensional.

With the extended models, we look for general properties that imply
dwell time symmetry, even if not all models with these properties are
biologically relevant. We find that the strong coupling and bottleneck
assumptions mentioned in the introduction are sufficient.  This means
that if dwell time symmetry is observed not to hold for a particular
system, then one can conclude that the strong coupling or bottleneck
properties are absent. To model such a system, one must go beyond the
extended models presented below, for example along the lines discussed
in Refs.~\citep{bustamante01,reimann02,juelicher97}.  The special
situation without strong coupling but with the bottleneck property is
illustrated below for an example model (Fig.~\ref{N2example2D}).

The extension of our results to detachments and observable substeps in
later subsections is directly motivated by experimental observations
of such events \citep{uemura04,nishiyama02,taniguchi05}.
\subsection*{The Class of Extended Models}\label{EXTmodels}
In this section, we describe a large class of extended models which
display dwell time symmetry in a little more detail, and indicate how
the dwell time symmetry is derived.  We use chemical kinetics, but
write a Master equation with arbitrary transition rates $w_{ij}$ from
state $j$ to $i$,
\begin{equation}\label{Meq}
  \partial_tq_i(t)=\sum_{j\ne i}w_{ij}q_j(t) -\sum_jw_{ji}q_i(t).
\end{equation}
Non-zero ``diagonal'' rates $w_{ii}$ are allowed, for example to
describe irreversible detachments of motors from their tracks
\citep{kolomeisky00,kolomeisky05}, or other events that can be
filtered out experimentally.

We also need to specify the condition of strong coupling.  One way is
to demand that the transition rates around any closed loop
$i_0,i_1,i_2,\ldots,i_m=i_0$ of transitions satisfy
\begin{equation}\label{micrev}
  \frac{w_{i_1i_0}w_{i_2i_1}\cdots
  w_{i_{m-1}i_m}}{w_{i_0i_1}w_{i_1i_2}\cdots w_{i_mi_{m-1}}}=1.
\end{equation}
This property is automatically fulfilled for sequential models, and
makes it possible to define a free energy $G_j$ for each state $j$, up
to an additive constant. From detailed balance, we have
$w_{ij}/w_{ji}=e^{-(G_i-G_j)/\kBT}$ for the free energy difference
along the transition $i\leftrightharpoons j$. Hence, \Eq{micrev} says
that the sum of free energy differences along any closed loop is zero,
which means that free energy dissipation that produces no net motion,
e.g., futile ATP hydrolysis, is ruled out. (Note that the
mechanochemical cycle, say, from a state $(k)$ to $(k+N)$, is not a
closed loop in our description.)  We also assume that no transition is
irreversible, i.e., \mbox{$w_{ij}\ne 0\;\; \Leftrightarrow\;\;
w_{ji}\ne 0$}, although the transition rates may be arbitrarily small.

To reflect the cyclic operation of motor proteins, the transition
rates are periodic with period $N$, and there is a well defined free
energy per period,
\begin{equation}\label{genDG}
  w_{j+N,i+N}=w_{ij},\quad G_{i+N}=G_i+\Delta G.
\end{equation}

\begin{figure}[b]\begin{center}
  \includegraphics{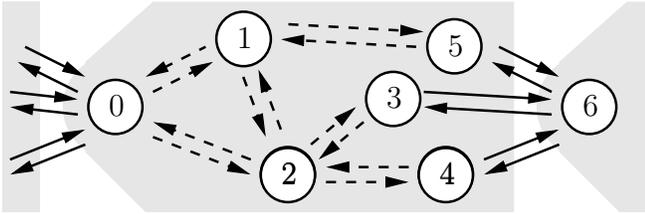}
\end{center}
  \caption{Example of a model with the bottleneck property and $N=6$
   states per cycle. All states inside the gray area belong to the
   same cycle. The bottleneck states are $0,\pm N,\pm 2N\ldots$. Black
   arrows denote transitions coupled to a mechanical step of length
   $d$, and dashed arrows indicate transitions within a cycle, which
   are undetectable. It is not possible to go from one cycle to the
   next without either leaving or arriving at a bottleneck state. In
   general, transitions between any states within a cycle are allowed,
   and bottleneck states might have a transition to any state in the
   neighbor cycle.  }\label{generalPeriodic}
\end{figure}

An important step in the derivation of \Eq{mainresult} is to conclude
that the state of the motor immediately after a step is independent of
earlier steps. In a sequential model, this occurs because there is
only one step-producing transition per cycle.  This condition can be
relaxed somewhat. It is sufficient to assume the bottleneck property
mentioned in the introduction, i.e., that all transitions
corresponding to a step are either to or from a single state, the
bottleneck state, as illustrated in \Fig{generalPeriodic}. As the term
indicates, the system must visit the bottleneck state each time it
goes through the cycle.

Whether the bottleneck assumption holds must be determined for each
system separately. For example, deviations from the step-step
correlations predicted in \Eq{kc} means that the assumption is not
valid.

Although the models considered in this section are more general than
the sequential ones, several properties of the sequential models are
retained. These include strong coupling, periodicity, and that steps
(roughly) correspond to one transition in the enzymatic cycle. They
therefore describe an effectively one-dimensional motion, where the
motion along the spatial and chemical reaction coordinates
\citep{bustamante01} are tightly coupled to each other for motion in
both directions. Hence, backward and forward motion proceeds in
opposite directions along the same reaction paths, just as for the
sequential models.
\subsection*{Dwell Time Symmetry and Detachments}
Dwell time symmetry for the extended models with detachments can be
derived using the same methods as for sequential models, and further
details on that derivation are given in appendix C.  Just as in the
sequential case, periodicity is only necessary to establish
\Eq{mainresult2}, while the dwell time symmetry, \Eq{mainresult1},
follows also without assuming periodicity.

Examples of the non-periodic case are the subcycles in
\Fig{substepExample}, which shows a model with two observable substeps
per cycle. Each step in the example satisfies the bottleneck property,
but both bottleneck states are in the same subcycle. Deriving dwell
time symmetry for the different subcycles proceeds as in appendix C,
with slightly different initial conditions.

As mentioned above, detachments of a motor from its track are observed
in several systems \citep{uemura04,nishiyama02,taniguchi05}.  We model
detachments by introducing death rates
\citep{kolomeisky05,kolomeisky00} as extra terms in the diagonal
elements of the Master equation and $\matris{M}$ matrix of
Eqs.~\eqref{seqMelements} and \eqref{genMelements}.

The presence of detachments affect the long time behavior of the model
\citep{kolomeisky00}, as well as dwell time distributions and pairwise
splitting probabilities.  Conditional detachment probabilities also
have to be added to Eqs.\ \eqref{splitNormalization} and
\eqref{stepMarkov}.

However, in the derivations of dwell time symmetry in appendix C, the
transition rates always enter as ratios of forward and reverse rates,
e.g., $M_{k_mk_{m+1}}/M_{k_{m+1}k_m}$ in \Eq{seqRatio}, so that
diagonal elements ($k_m=k_{m+1}$) always cancel. Hence,
\Eq{mainresult} remains valid even in the presence of detachments.

\subsection*{Observable Substeps}\label{detsub}
\begin{figure}[b]\begin{center}
  \includegraphics{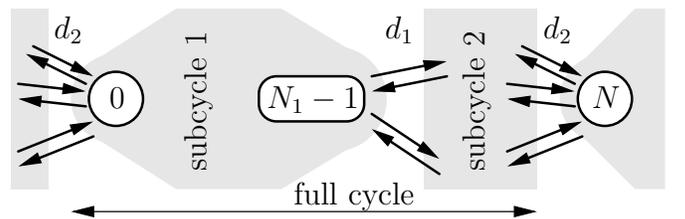}
\end{center}
  \caption{Simple periodic model with two substeps per cycle, with
   step lengths $d_1$ and $d_2$. Subcycle 1 has $N_1$ states, and
   subcycle 2 has $N-N_1$ states. Both steps have the bottleneck
   property, although the bottleneck states $0$ and $N_1-1$ are both in
   subcycle 1. If strong coupling is satisfied, both subcycles
   display dwell time symmetry.
  }\label{substepExample}
\end{figure}
For some motors, the full forward and backward steps are divided into
observable substeps \citep{uemura04,yasuda01}. In this section, we
show how the relation between splitting probabilities in
\Eq{mainresult} can be generalized, if strong coupling holds and all
substeps satisfy the bottleneck property.

To start with, consider the system in \Fig{substepExample}, which
produces substeps $d_1$ and $d_2=d-d_1$ during each cycle, and assume
that each sub-cycle displays dwell time symmetry. From now on, we
denote a forward step of length $d_1$ with $+_1$, and similarly for
the other substeps.  Keeping track of only the $d_2$-steps, we can
analyze the system as in previous sections, using four splitting
probabilities $\pi'_{\pm_2\pm_2}=\pi_{\pm\pm}$ and dwell times
$\tau'_{\pm_2\pm_2}(t)=\tau_{\pm\pm}(t)$, which satisfy
\Eq{mainresult}.

On the other hand, if we keep track of and discriminate between all
substeps, we could instead measure eight splitting probabilities:
$\pi_{+_1+_2}$, $\pi_{+_1-_1}$, $\pi_{-_1+_1}$, $\pi_{-_1-_2}$, and so
on. Summing over all paths between two consecutive $+_2$-steps, we get
\begin{multline}\label{f2f2}
    \pi'_{+_2+_2}=\pi_{+_2+_1}\Big(1+\pi_{+_1-_1}\pi_{-_1+_1}+
    \ldots\Big)\pi_{+_1+_2}\\
    =\pi_{+_2+_1}\sum_{k=0}^\infty(\pi_{+_1-_1}\pi_{-_1+_1})^k\pi_{+_1+_2}
    =\frac{\pi_{+_2+_1}\pi_{+_1+_2}}{1-\pi_{+_1-_1}\pi_{-_1+_1}},
\end{multline}
and similarly
\begin{equation}\label{b2b2}
    \pi'_{-_2-_2}=\frac{\pi_{-_2-_1}\pi_{-_1-_2}}{1-\pi_{+_1-_1}\pi_{-_1+_1}}.
\end{equation}
This means that the relation between free energy and splitting
probabilities for a system with only one step per cycle generalizes to
\begin{equation}\label{splitRatio2}
  \frac{\pi_{+_2+_1}\pi_{+_1+_2}}{\pi_{-_2-_1}\pi_{-_1-_2}}=
  \frac{\pi'_{+_2+_2}}{\pi'_{-_2-_2}}=
  e^{-\Delta G/\kBT}.
\end{equation}
Note that we summed over all paths between two $+_2$ steps without
explicit reference to all paths that start with a $+_2$ step and end
with something else. This means that \Eq{splitRatio2} is valid also
for a system with detachments, or with several parallel pathways.

For systems which go through substeps $d_1$, $d_2$ \ldots $d_K$ in
each cycle, we can use the same argument to relate the splitting
probabilities of the full analysis (all substeps included) to one
where the first substep ($d_1$) is ignored:
\begin{align}
    \pi_{+_K+_2}'&=\frac{\pi_{+_K+_1}\pi_{+_1+_2}}
       {1-\pi_{+_1-_1}\pi_{-_1+_1}},\\
    \pi_{-_2-_K}'&=\frac{\pi_{-_2-_1}\pi_{-_1-_K}}
       {1-\pi_{+1-_1}\pi_{-_1+_1}}.   
\end{align}
Iterating this transformation to ignore substeps $2,3,\ldots,K-1$ as
well, we find the following relation for a cycle with $K$ visible
substeps:
\begin{equation}\label{splitProd}
  \frac{\pi_{+_K+_1}}{\pi_{-_1-_K}}
  \frac{\pi_{+_1+_2}}{\pi_{-_2-_1}}\cdots
  \frac{\pi_{+_{(K-1)}+_K}}{\pi_{-_K-_{(K-1)}}}=
  \frac{\pi'_{+_K+_K}}{\pi'_{-_K-_K}}=
  e^{-\Delta G/\kBT}.
\end{equation}
In a complicated system with parallel pathways where a cycle can be
completed using different sequences of substeps, \Eq{splitProd} holds
for every such sequence separately.
\subsection*{An Example Model}\label{examplemodel}
\begin{figure}[b]\begin{center}
    \includegraphics{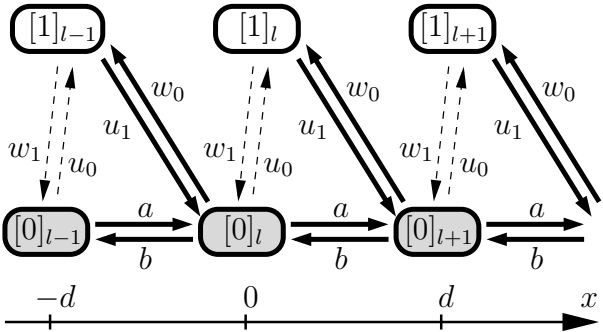}
  \end{center}
  \caption{Example of a non-sequential model with $N=2$ states per
   cycle, with two parallel pathways that produce a step in the $x$
   direction. Solid arrows are step-producing transitions, while
   dashed arrows indicate hidden transitions that do not produce an
   observable step. Shaded boxes are bottleneck states.
   }\label{N2example1D}
\end{figure}
In this section, we consider a small but non-trivial model with 2
states per cycle, to present an explicit example that illustrates the
results of the previous sections.

The model has two parallel pathways as sketched in \Fig{N2example1D},
and the steady state velocity and effective dispersion are known
exactly \citep{kolomeisky01}. The bottleneck property is satisfied,
and strong coupling is equivalent to 
\begin{equation}\label{N2rev}
  u_0u_1a=w_0w_1b \Leftrightarrow
  \frac{u_0u_1}{w_0w_1}=\frac{a}{b}\equiv e^{-\Delta G/\kBT}.
\end{equation}

Computing the dwell time distributions and splitting probabilities
(see appendix D for details), we get
\begin{align}\label{ffbbExample}
  \piff&=\big( a(u_1+w_1)+u_0u_1\big)/c_0,\\ \pibb&=\big(
  b(u_1+w_1)+w_0w_1\big)/c_0,\\
  \rff(t)&=c_0\frac{e^{\lambda_1t}-e^{\lambda_2t}}{\lambda_1-\lambda_2}
  +\frac{a}{\piff}\frac{\lambda_1e^{\lambda_1t}-\lambda_2e^{\lambda_2t}}
  {\lambda_1-\lambda_2},\\
  \rbb(t)&=c_0\frac{e^{\lambda_1t}-e^{\lambda_2t}}{\lambda_1-\lambda_2}
  +\frac{b}{\pibb}\frac{\lambda_1e^{\lambda_1t}-\lambda_2e^{\lambda_2t}}
  {\lambda_1-\lambda_2}.
\end{align}
The eigenvalues $\lambda_{1,2}$ are given in appendix D, and
$c_0=\lambda_1\lambda_2>0$.  For this model, \Eq{N2rev} is obviously
equivalent to the dwell time symmetry of \Eq{mainresult}.  For the
other two pairs of steps, we get
  \begin{align}
    \pifb&=(b+w_0)(u_1+w_1)/c_0,\\
    \pibf&=\big(u_1(u_0+w_0)+a(u_1+w_1)\big)/c_0,\\
    \rfb(t)&=c_0\frac{e^{\lambda_1t}-e^{\lambda_2t}}{\lambda_1-\lambda_2}
    +\frac{b+w_0}{\pifb}\,
    \frac{\lambda_1e^{\lambda_1t}-\lambda_2e^{\lambda_2t}}
    {\lambda_1-\lambda_2},\\
    \rbf(t)&=c_0\frac{e^{\lambda_1t}-e^{\lambda_2t}}{\lambda_1-\lambda_2}
    +\frac{ab+w_0u_1}{\pibf(b+w_0)}\,
    \frac{\lambda_1e^{\lambda_1t}-\lambda_2e^{\lambda_2t}}
    {\lambda_1-\lambda_2}.\label{fbbfExample}
  \end{align}
We see that $\rfb(t)$ and $\rbf(t)$ differ both from each
other and from the $++$ and $--$ distributions, independent of whether
\Eq{N2rev} holds or not. The step directions are anti-correlated,
since
\begin{equation}\label{N2Xgamma}
  \gamma=\piff+\pibb-1=-w_0u_1/c_0<0.
\end{equation}
\begin{figure}[b]\begin{center}
    \includegraphics{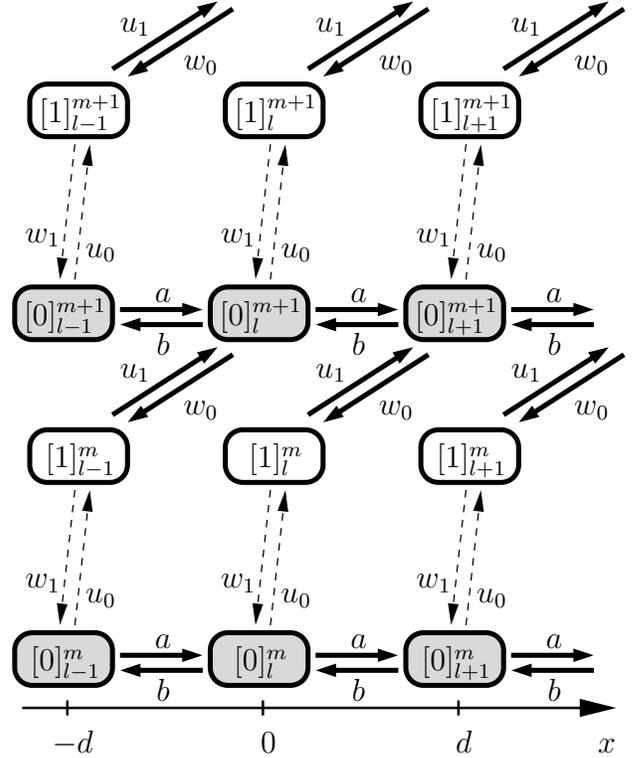}
  \end{center}
  \caption{A model with one spatial and one chemical reaction
   coordinate, denoted by indices $l$ and $m$ respectively. Shaded
   boxes are bottleneck states.  Projecting out the chemical
   coordinate as explained in the text, this model can be brought to
   the same form as in \Fig{N2example1D}. In general, the resulting
   effective one-dimensional model does not satisfy the strong
   coupling assumption of \Eq{N2rev}.
  }\label{N2example2D}
\end{figure}

The example illustrates the logic of our results. Strong coupling and
the bottleneck property together are sufficient conditions for the
dwell time symmetry.  The bottleneck property alone is sufficient for
the Markov-chain description of step directions in \Eq{stepMarkov},
which remains valid even if \Eq{N2rev} is not satisfied.

A possible reason for why the model in \Fig{N2example1D} might not
satisfy \Eq{N2rev} is illustrated in \Fig{N2example2D}. This is again
an $N=2$ state model, but with an independent ``chemical'' reaction
coordinate (the superscript $m$ on the states), e.g, the number of
hydrolyzed ATP molecules \citep{bustamante01}.  For example, the path
$[0]_l^m\leftrightharpoons [1]_l^m\leftrightharpoons [0]_{l+1}^{m+1}$
could be a step driven ATP hydrolysis, while the reaction
$[0]_l^m\leftrightharpoons [0]_{l+1}^m$ could describe an ATP
independent step.

This model can be transformed to the 1D model in \Fig{N2example1D}
through the projection $[j]_l=\sum_m [j]_l^m$, which
leaves the rates unaffected. However, the open-ended reaction path
$[0]_l^m\leftrightharpoons [1]_l^m\leftrightharpoons
[0]_{l+1}^{m+1}\leftrightharpoons [0]_l^{m+1}$ is transformed into the
closed loop $[0]_l\leftrightharpoons[1]_l\leftrightharpoons
[0]_{l+1}\leftrightharpoons [0]_l$. Since strong coupling places no
constraints on the rates in the original model in \Fig{N2example2D},
there is no reason to assume that \Eq{N2rev} holds for the projected
model.  In the effective 1D description, this loop then becomes a slip
loop, which dissipates free energy without producing net
motion. Hence, the simple step-step correlations of \Eq{kc} is
retained, but not the dwell time symmetry.

In systems where it is not possible to measure the ``chemical
position'' with single molecule precision, the resulting violation of
dwell time symmetry is a useful test for such projected slip loops, if
the bottle-neck property is satisfied.  This property can in turn be
ruled out, for example if the step-step correlations are more
complicated than predicted in \Eq{kc}.

The most interesting candidates for such tests seem to be the rotary
motors driven by ion fluxes, e.g., the F$_\text{O}$ part of ATP
synthase, or the bacterial flagellar motor, since it is difficult to
measure the ion flow with single molecule precision. We are not aware
of any experiments on ATP synthase under conditions that produces both
forward and backward steps. Data from a flagellar motor will be
analyzed in the next section.
\section*{Application to the Flagellar Motor}\label{flagellarSteps}
\begin{table}[b]
  $$\begin{array}{crrrrrrrrr}
  \hline
    ij&+&-&\delta&{++}&{+-}&+\delta&{-+}&{--}&-\delta\\ 
    \hline
    \multicolumn{10}{l}{\text{Trajectory A:}}\\ 
    \hline 
    n_{ij} & 108 & 103 & 13 & 30 & 68 & 10 & 69 & 30 & 3\\
    \pi_{ij}&0.48 & 0.46 & 0.06 & 0.28 & 0.63 & 0.09 & 0.68 & 0.29 &
    0.03\\ 
    \hline
    \multicolumn{10}{l}{\text{\rule{0pt}{0pt}Trajectory B:}}\\ 
    \hline
    n_{ij} & 43 & 27 & 11 & 20 & 16 & 6 & 17 & 5 & 5 \\ 
    \pi_{ij}& 0.53 & 0.33 & 0.14 & 0.48 & 0.38 & 0.14 & 0.63 & 0.19 &
    0.19\\
  \end{array}$$
  \caption{Steps ($\pm$) and drift ($\delta$) events for the two
     analyzed stepping trajectories.  A forward step followed by a
     drift step is denoted $+\delta$. The rows display the number of
     events ($n_{ij}$) and the corresponding conditional probabilities
     ($\pi_{ij}$). The average velocities and average dwell times are
     zero and $0.04$ s for trajectory A, and $1.1$ Hz and $0.007$ s
     for trajectory B.}\label{stepDir}
\end{table}
In this section, we apply our theoretical results to analyze stepping
data from a recent experiment with a chimaeric flagellar motor
\citep{sowa05}.  As we will see, the data is consistent with the
predicted dwell time symmetry and step-step correlations within the
experimental uncertainty. We also estimate the free energy per cycle,
and show that the estimate based on cycle completions can have a
significant systematic error, compared to \Eq{mainresult2}.

The flagellar motor propels many swimming bacteria, by driving the
rotation of flagellar filaments. Each filament is driven at its base
by a transmembrane rotary motor, powered by ion flux (Na$^+$ in this
case \citep{sowa05}) down an electrochemical gradient across the cell
membrane. The motor is about 45 nm in diameter, and is believed to
contain 13 torque-generating units \citep{sowa05,berg03}. In this
experiment, only one unit was active, and the flagellum is expected to
switch between 26 distinct orientations per turn, corresponding to a
step length of $360^\circ/26\approx 14^\circ$ \citep{sowa05}. We
analyzed one trajectory with close to zero net velocity, and one with
finite velocity.

\begin{figure}[b]\begin{center}
    \includegraphics{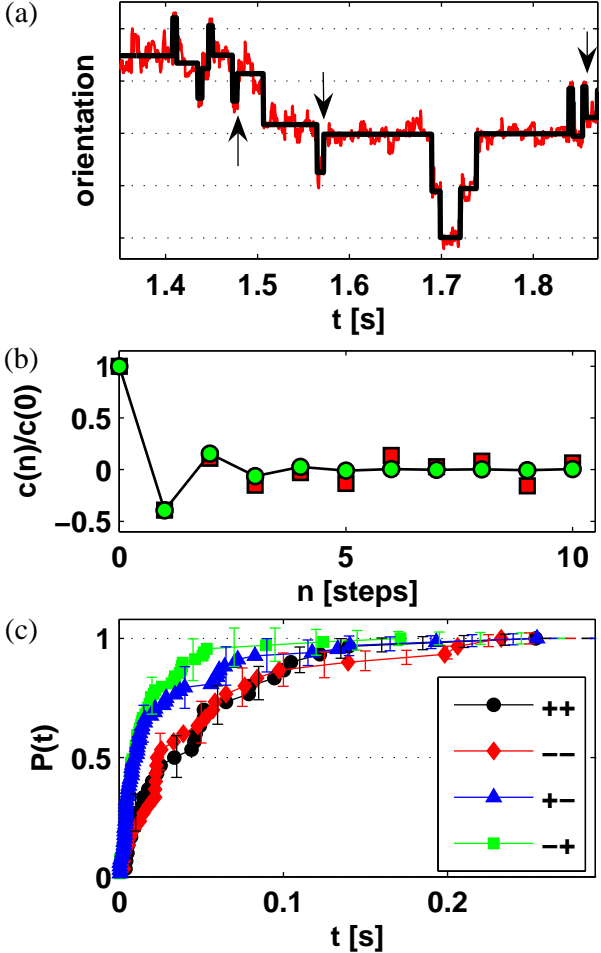}
  \end{center}
  \caption{Analysis of stepping data from trajectory A. (a) Stepping
   data and result of step detection (staircase line).  Grid lines
   indicate the theoretical step length $d=13.8^\circ$. Stepping is
   consistent with the theoretical step length in intervals
   interrupted by drift events (arrows), and adjacent intervals are
   out of phase with each other. (b) Step-step autocorrelation
   function as defined in \Eq{kc}. The measured correlation function
   (squares) is consistent with the theoretical prediction (circles)
   using conditional splitting probabilities from
   Tab.~\ref{stepDir}. Symbols are the same size as the estimated
   standard deviations.  (c) Distribution functions $P_{\pm\pm}(t)$
   for the four different dwell times.  }\label{flagellaPC}
\end{figure}

Aided by Chung-Kennedy filtering \citep{chung91} to enhance the steps,
and the step-finding algorithm described in Ref.~\citep{sowa05}, the
raw data was converted to stair-case stepping data, as shown in
\Fig{flagellaPC}(a). Upon close inspection, the trajectory seems to be
divided into intervals where the step lengths are consistent with each
other, but out of phase with steps outside. This apparent drift might
be due to dynamical exchange of the stator units \citep{reid06} that
anchor the motor to the cell.

We identified intervals of consistent stepping by inspection (see
\Fig{flagellaPC}(a)), and excluded the drift from our analysis by
treating inconsistent steps as detachments, i.e., each interval of
consistent stepping was treated as an independent run.  The result is
summarized in Tab.~\ref{stepDir}, and we now proceed to compare
step-step correlations and dwell time distributions from trajectory A
with the theoretical predictions. Trajectory B contained too few steps
to make such comparisons meaningful.

The step-step autocorrelation function for trajectory A is shown in
\Fig{flagellaPC}(b).  The prediction $C(n)/C(0)=\gamma^n$ of
\Eq{stepMarkov} is valid for trajectories without detachments. In
order to compare theory and experiments, the splitting probabilities
in Tab.~\ref{stepDir} must therefore be renormalized by a factor
$(1-\pi_{\pm\delta})^{-1}$ to account for the fact that intervals of
consistent stepping (by definition) contain no drift events. This
gives the theoretical prediction
\begin{equation}\label{gammaPrediction}
\gamma=\frac{\piff}{1-\pi_{+\delta}}+\frac{\pibb}{1-\pi_{-\delta}}-1=
-0.43\pm0.06.
\end{equation}
A least squares fit to the experimental correlations gives
$\gamma=-0.40$, in agreement with \Eq{gammaPrediction}, but
inconsistent with uncorrelated step directions.

Having confirmed that the step-step correlations are consistent with
\Eq{kc}, we go on to test the dwell time symmetry. The empirical
distribution functions $P_{\pm\pm}(t)$ are plotted in
\Fig{flagellaPC}(c) for trajectory A. Error bars are standard
deviations from bootstrap estimates \cite{numres}.  Using the
Kolmogorov-Smirnov test \citep{barlow}, we conclude, with 95\%
confidence, that $\tbf$ have different distribution than $\tff$ and
$\tbb$, and that $\tfb$ have different distribution than
$\tbb$. However, the test could not detect significant differences
between the other three pairs ($++/--$, $+-/--$, and $+-/-+$). This is
again consistent with the theoretical predictions, and also shows that
the statistics is good enough to detect differences between
distributions that are not equal.  It is also interesting to note from
\Fig{flagellaPC}(c) that $\tff$ and $\tbb$ are longer than $\tfb$ and
$\tbf$ on average. This is reasonable, since $+-/-+$ events in
principle only requires a single transition, while the system must go
through a complete cycle to complete a $++$ or $--$ event.

The ionic motive force was not measured independently \citep{sowa05},
so it is not possible to compare the free-energy estimate of
\Eq{mainresult} with an experimental value. However, we can compare
our result with that based on \Eq{cycleresult}, which was used in
Ref.~\citep{sowa05}. The results are summarized in Tab.\
\ref{dGestimates}.  As expected, the free energy per cycle is close to
zero in trajectory A, which has almost no net velocity, while
trajectory B clearly has a finite free energy per cycle to drive the
rotation. Also note the differences in the estimates based on
\Eq{mainresult} and \Eq{cycleresult}, which show that the
cycle-completion estimate can give rise to significant systematic
errors in estimated free energies if applied to stepping data.
\begin{table}[b]
  $$\begin{array}{cr@{\pm}lr@{\pm}l}
    &\multicolumn{2}{c}{\Delta G_\text{steps}}& 
    \multicolumn{2}{c}{\Delta G_\text{cycles}}\\
    \hline
    \text{Trajectory A}&  0.06 & 0.23 & -0.05 & 0.10\\
    \text{Trajectory B}& -0.94 & 0.28 & -0.47 & 0.13\\    
    \hline
  \end{array}$$
  \caption{Free energy estimates for trajectories A and B, in units of
    $\kBT$. The estimate $\Delta G_\text{steps}$ is based on our main
    result, \Eq{mainresult}, while $\Delta G_\text{cycles}$ comes from
    \Eq{cycleresult}, i.e., treating steps as cycle
    completions.}\label{dGestimates}
\end{table}
\section*{Conclusion and Outlook}\label{conclusion}
The statistical properties of steps in reversible molecular motors are
not the same as that of cycle completions in enzyme kinetics. To
interpret stepping trajectories correctly, this difference must be
taken into account.

In this paper, we have extended the theory for dwell times in
sequential models \citep{tsygankov06a,qian97}, and derived a symmetry
relation for the conditional dwell time distributions for a large
class of experimentally relevant models, including parallel pathways,
visible substeps, and detachments. In contrast to the statistics of
cycle completions \citep{kolomeisky05,qian06,wang06}, the dwell time
distributions and splitting probabilities of steps depend on the step
directions of both steps before and after the dwell period.

The dwell time symmetry is a consequence of strong coupling, i.e.,
tight coupling for both forward and backward steps, and a bottleneck
property of the underlying kinetic scheme. This means that the motion
of the system is essentially a one-dimensional random walk in the
(two-dimensional) space of spatial and chemical reaction coordinates.
This makes it possible to use our results to infer information about
the coupling and efficiency of a system from kinetic data, without
detailed assumptions about the underlying kinetic scheme. In this
respect, it is similar to the relation $r\ge 1/N$ between the
randomness parameter $r$ and the number of steps $N$ in the
mechanochemical cycle of molecular motors
\citep{koza02,fisher99a,svoboda94}.

An alternative to the discrete formalism used here, is to include
continuous spatial degrees of freedom, as is common for example when
modeling ratchet-type motor mechanisms
\cite{juelicher97,bustamante01,reimann02}. Since such models can be
discretized \cite{wang03}, it is in principle possible to inquire
about strong coupling and the bottleneck property in continuous models
as well. An example of this is diffusion in one dimension (see
Fig.~\ref{contLim}), which strictly satisfies strong coupling and the
bottleneck property. An interesting problem for further research is to
quantify how well the dwell time symmetry is preserved in system with
small deviations from these assumptions.

We analyzed stepping data from single motor subunits in a flagellar
motor \citep{sowa05}. The data seems to be consistent with the
predicted dwell time symmetry and step-step correlations, although
only one trajectory with almost zero velocity had enough steps to make
such a comparison meaningful. 

The form of the step-step correlations is consistent with an
underlying kinetic scheme that satisfies the bottleneck property. So,
does the dwell time symmetry in Fig.~\ref{flagellaPC} indicate that
the flagellar motor subunits are strongly coupled to the driving ion
flow?  Another possibility is that the steps are equilibrium
fluctuations, as indicated by the low $\Delta G$ estimated from the
stepping probabilities. In equilibrium, there is no free energy change
associated with ion transport, which means that dwell time symmetry
could be obtained also for a system with loose coupling.  To say
anything about the coupling in the flagellar motor, the dwell time
symmetry must be tested in a regime with finite (and constant)
velocity and ionic driving force. This is in principle a question of
observing more steps in such conditions, but a systematic way to
identify and separate steps from drift events would also be useful.

The step-step correlations predicted in \Eq{kc} might be present also
in systems that violates the dwell time symmetry. Deviations from this
form would say something about the topology of the kinetic pathways in
such systems, and it would probably be useful to apply our analysis
also to kinesin and myosin V.

An accurate theoretical model is often crucial in order to correctly
interpret experiments on such complex systems as motor proteins. Many
previous theoretical works on steps and dwell times in molecular
motors \citep{kolomeisky05,kolomeisky03,shaevitz05,santos05} derive or
assume descriptions where consecutive steps or cycles are
statistically independent of each other. As we have demonstrated, this
does not apply to the flagellar motor of Ref.~\citep{sowa05}.  We have
also presented a large class of simple models where easily accessible
quantities like the dwell times and step directions are correlated.

We expect our results to be of practical use in both data analysis and
design of experiments.  In particular, further experimental efforts
are motivated in order to detect correlations and collect significant
statistics for both forward and backward steps.

\begin{appendix}
\section*{Appendix A}\label{seqDwellTimeSymmetry} 
We now give a detailed derivation of \Eq{mainresult} for a sequential
model.  The initial condition is the state immediately after a step,
and can be written
\begin{equation}\label{seqIC}\begin{split}    
    q_j(0)&=q^{(+)}_j=\delta_{j,0}\text{, after a forward step, and}\\
    q_j(0)&=q^{(-)}_j=\delta_{j,N-1}\text{, after a backward step}.
\end{split}\end{equation}
To derive \Eq{mainresult}, we compute $\piff\partial_t^{\,n}P_{++}(0)$
and $\pibb\partial_t^{\,n}P_{--}(0)$ using \Eq{seqPFdef} and the initial
conditions in \Eq{seqIC}. This gives
\begin{multline}\label{seqFFdiff}
    \piff\partial_t^{\,n}\,P_{++}(0)=
  u_{N-1}\big(\matris{M}^{n-1}\vektor{q}^{(+)}\big)_{N-1}\\
  =u_{N-1}\sum_{\{k_j\}}M_{N-1,k_{n-2}}\cdots M_{k_2,k_1}M_{k_1,0}\,,
  \end{multline}
where the summation over $k_1,k_2,\ldots$ goes from $0$ to $N-1$.  The
same calculation for $P_{--}(t)$ yields
\begin{multline}\label{seqBBdiff}
\pibb\partial_t^{\,n}\,P_{--}(0)=
w_0\big(\matris{M}^{n-1}\vektor{q}^{(-)}\big)_0=\\
  =\sum_{\{k_j\}}M_{0k_1}M_{k_1,k_2}\cdots M_{k_{n-2},N-1}w_0\,.
\end{multline}
The products of matrix elements in Eqs.\
(\ref{seqFFdiff}-\ref{seqBBdiff}) correspond to $n$-step paths between
states $0$ and $N-1$, plus the extra escape step.  For sequential
models, the shortest such path is $N$ steps, so the first $N-1$
derivatives are zero.  For $n\ge N$, we note that there is a
one-to-one correspondence between the nonzero terms in \Eq{seqFFdiff}
and \Eq{seqBBdiff}. For each term including a path from $0$ to $N-1$
in \Eq{seqFFdiff}, there is a corresponding term for the reverse path
from $N-1$ to $0$ in \Eq{seqBBdiff}.  The ratio of two corresponding
non-zero terms is
\begin{equation}\label{seqRatio}
  R(\{k_j\})=
       \frac{M_{k_1,0}}{M_{0,k_1}}\frac{M_{k_2,k_1}}{M_{k_1,k_2}}
       \cdots
       \frac{M_{N-1,k_{n-2}}}{M_{k_{n-2},N-1}}\frac{u_{N-1}}{w_0}.
\end{equation}
Going back to \Eq{seqMelements} for the elements of $\matris{M}$, and
using $w_0=w_{N}$ by periodicity, we see that $R(\{k_j\})=e^{-\Delta
G/\kBT}$ for all pairs of corresponding terms.  Moreover,
$P_{++}(0)=P_{--}(0)=0$, as there is no transition which produces two
steps at once. Therefore,
\begin{equation}\label{seqTaylor}
  \piff\partial_t^{\,n}P_{++}(0)=
  e^{-\Delta G/\kBT}\pibb\partial_t^{\,n}P_{--}(0)
\end{equation}
for all $n\ge 0$. The underlying reason is that the sum of free energy
changes is the same along all possible paths going forward one cycle
from $0$. 

Since $P_{++}(t)$ and $P_{--}(t)$ are integrals of the solutions of
the reduced Master equation, which is a finite system of ordinary
differential equations with constant coefficients, they are smooth
functions which have Taylor series. Hence, \Eq{seqTaylor} together
with the normalization in \Eq{Pnorm} implies \Eq{mainresult}.  A
similar correspondence between paths was used in Ref.~\citep{wang06}
to derive \Eq{cycleresult}. The periodicity is necessary to get the
simple relationship between $\piff$, $\pibb$ and $\Delta G$ in
\Eq{mainresult2}, but not to establish the dwell time symmetry in
\Eq{mainresult1}. Without the periodicity, \Eq{mainresult2} is
replaced by
\begin{equation}
  \frac{\piff}{\pibb}=R(\{k_j\})=\frac{u_{N-1}}{w_0}e^{-(G_{N-1}-G_0)/\kBT}.
\end{equation}

\section*{Appendix B}\label{seqPDF}
To derive explicit expressions for $\rho_{++}(t)$ and $\rho_{--}(t)$
for sequential models, we first note that the occupation probabilities
$q_j(t)$ in \Eq{seqPFdef} are the solution of a system of linear
ordinary differential equations with constant coefficients, namely the
elements of $\matris{M}$ in \Eq{seqMelements}.  If the eigenvalues
$\{\lambda_j\}$ of $\matris{M}$ are non-degenerate, the solutions have
the form $q_j(t)=\sum_{i=1}^Na_i^{(j)}e^{\lambda_it}$.  Note that the
eigenvalues have negative real parts, to guarantee that the system
eventually leaves the interval. We make the ansatz
\begin{equation}\label{seqAlphaAnsatz}
  P_{++}(t)= P_{--}(t)=1+\sum_{j=1}^N \alpha_je^{\lambda_jt},
\end{equation}
where the term 1 ensures proper normalization.  We need $N$ equations
to determine the coefficients $\alpha_j$. These are
\begin{equation}
  \partial_t^{\,k}P_{++}(0)=\partial_t^{\,k}P_{--}(0)=0 
\end{equation}
for $0\le k\le N-1$.  The case $k=0$ was argued just before
\Eq{seqTaylor}, and $1\le k\le N-1$ follows from the argument between
Eqs.~\eqref{seqBBdiff} and \eqref{seqRatio}. The resulting system of
equations for $\alpha_j$ is of the Vandermonde type,
\begin{equation}\label{seqAlphaEq}
  \left(\begin{array}{cccc} 1&1&\ldots&1\\
    \lambda_{1}&\lambda_{2}&\ldots&\lambda_{N}\\
    \lambda_{1}^2&\lambda_{2}^2&\ldots&\lambda_{N}^2\\ 
    \vdots&\vdots &    \ddots&\vdots\\
    \lambda_{1}^{N-1}&\lambda_{2}^{N-1}&\ldots&
    \lambda_{N}^{N-1}\\
  \end{array}\right)
  \left(\begin{array}{c}\alpha_1\\\alpha_2\\\alpha_3\\
    \vdots\\\alpha_N\end{array}\right)
    =
  \left(\begin{array}{c}-1\\0\\0\\\vdots\\0\end{array}\right).
\end{equation}
Solving with Cramer's rule and setting
\begin{equation}
  \rff(t)=\rbb(t)=\partial_t P_{--}(t)=
  \sum_{k=1}^{N}\lambda_k\alpha_ke^{\lambda_kt},
\end{equation}
we arrive at \Eq{seqdist}. The distributions of $\tfb$ and $\tfb$ can
be computed with the same ansatz, but the derivatives
$\partial_t^{\,k}P_{+-}(0)$ and $\partial_t^{\,k}P_{+-}(0)$ are
non-zero for $k\ge 1$, so the results are more complicated and there
is in general no simple relation between $\rfb(t)$ and $\rbf(t)$.

\section*{Appendix C}\label{extendedDTS}
In this appendix, we derive \Eq{mainresult} for a periodic extended
model like the one in \Fig{generalPeriodic}, generalizing the first
passage problem as we go along.  The derivation proceeds much as for
the sequential model, by solving a reduced Master equation for the
states $0,1,\ldots,N-1$,
\begin{equation}\label{genMelements}
  \partial_t\vektor{q}=\matris{M}\vektor{q},\quad
  M_{ij}=w_{ij}-\delta_{ij}\sum_{k=-\infty}^{\infty} w_{ki},
\end{equation}
with absorbing boundaries at $-1$ and $N$, and allowing $w_{ii}\ge 0$
to model detachments.  Note that the sum in \Eq{genMelements} have at
most $2N-1$ terms, due to the bottleneck property.  After a forward
step, the system is in state $0$, while it could in principle be
anywhere in the cycle after a backward step. The initial conditions,
describing the distribution of states just after a $\pm$ step, are
therefore
\begin{equation}\label{genIC}
    q_j(0)=q^{(+)}_j=\delta_{j,0},\quad
    q_j(0)=q^{(-)}_j=\frac{w_{jN}}{z_-},\quad
    z_-=\sum_{k=0}^{N-1}w_{kN}.
\end{equation}
Since there are several escape transitions for each step,
\Eq{seqPFdef} for the dwell time probability functions is generalized
to
\begin{equation}\label{genPFdef}
  \begin{split}
  \pi_{\pm+}P_{\pm+}(t)&=\sum_{k=0}^{N-1}w_{Nk}\int_0^tq_{k}(t)\du t\\
  \pi_{\pm-}P_{\pm-}(t)&=z_-'\int_0^tq_{0}(t)\du t,
  \quad z_-'=\sum_{k=-N}^{-1}w_{k0}.
  \end{split}
\end{equation}
We now proceed to compute derivatives of $P_{++}(0)$ and $P_{--}(0)$,
and note that by Eqs.~(\ref{micrev}-\ref{genDG}), the free energy
difference $G_N-G_0=\Delta G$ is independent of path:
\begin{multline}
  \piff\partial_t^{\,n}\,P_{++}(0)=\sum_{k_{n-1}=0}^{N-1}w_{Nk_{n-1}}
  \big(\matris{M}^{n-1}\vektor{q}^{(+)}\big)_{k_{n-1}}\\
  =\sum_{\{k_j\}}w_{Nk_{n-1}}M_{k_{n-1}k_{n-2}}\cdots M_{k_1k_0}
  \delta_{k_0,0}\,,
\end{multline}
\begin{multline}
  \pibb\partial_t^{\,n}\,P_{--}(0)=
  z_-\big(\matris{M}^{n-1}\vektor{q}^{(-)}\big)_0\\
  =z_-'\sum_{\{k_j\}}M_{0k_1}M_{k_1k_2}\ldots M_{k_{n-2}k_{n-1}}
  \frac{w_{k_{n-1}N}}{z_-}\\
  =\frac{z_-'}{z_-}e^{(G_N-G_0)/\kBT}\partial_t^{\,n}\,P_{++}(0).
\end{multline}
From the periodicity of $w_{ij}$ expressed in \Eq{genDG}, we get
$z_-'=z_-$, which proves \Eq{seqTaylor} and thereby
\Eq{mainresult}. The derivation for the case when backward steps
(instead of forward steps) end in the bottleneck state is analogous.

Note that our definition of the bottleneck property means that there
are no transitions that produce two steps at the same time. This rules
out discontinuities at $t=0$ in the distribution functions
$P_{\pm\pm}(t)$, and justifies our use of Taylor expansions.

\section*{Appendix D}\label{exApp}
In this appendix, we solve the example model in Fig.~\ref{N2example1D}
in some detail. In addition to deriving
Eqs.~(\ref{ffbbExample}-\ref{N2Xgamma}), this also illustrates a
method that can easily be generalized to larger systems and
implemented on computer algebra systems.  For $N\ge 5$ states per
cycle, the eigenvalues $\lambda_j$ must be calculated
numerically. With this exception, the method gives explicit
expressions for the dwell time distributions in the real time domain,
that can be directly compared to experimental dwell time histograms.

The first escape problem for the model in \Fig{N2example1D} is
governed by a reduced Master equation as described in
\Eq{genMelements}, with the matrix 
\begin{equation}
  \matris{M}=\left(\begin{array}{cc}
    -(a+b+u_0+w_0) & w_1\\
    u_0 & -(u_1+w_1)
  \end{array}\right).
\end{equation}
The eigenvalues of $\matris{M}$ are
\begin{equation}
    \lambda_{1,2}=\frac{-c_1\pm\sqrt{(w_0+u_0-w_1-u_1+a+b)^2+4u_0w_1}}{2},
\end{equation}
and satisfy the relations
\begin{align}
    c_0&=\lambda_1\lambda_2=(w_0+a+b)(u_1+w_1)+u_0u_1,\\
    c_1&=-(\lambda_1+\lambda_2)=a+b+u_0+w_0+u_1+w_1.
\end{align}

The initial conditions after forward and backward steps are given by
\Eq{genIC}, and the conditional dwell time distributions are given by
\Eq{genPFdef}: 
\begin{align}
  q_j^{(+)}(0)&=\delta_{0,j},\quad
  q_j^{(-)}(0)=\frac{b\delta_{0,j}+w_0\delta_{1,j}}{b+w_0},\label{Aic}\\
    \pi_{\pm-}P(t)_{\pm-}&=(w_0+b)\int_0^t q_0(t')dt',\\
    \pi_{\pm+}P(t)_{\pm+}&=\int_0^t aq_0(t')+u_1q_1(t')dt'.\label{Apdfdef}
\end{align}
Using the ansatz in \Eq{seqAlphaAnsatz} together with Eqs.\
(\ref{Aic}-\ref{Apdfdef}) to compute $\partial_t \pi_{\pm-}
P_{\pm-}(t)$ and $\partial_t \pi_{\pm+}P_{\pm+}(t)$, we get the
following systems of linear equations:
\begin{multline}
  \left(\begin{array}{cc}
    1&1\\ \lambda_1&\lambda_2
    \end{array}\right)
  \left(\begin{array}{cccc}
    \alpha_{1}^{++}&\alpha_{1}^{+-}&\alpha_{1}^{-+}&\alpha_{1}^{--}\\
    \alpha_{2}^{++}&\alpha_{2}^{+-}&\alpha_{2}^{-+}&\alpha_{2}^{--}\\
    \end{array}\right)\\
  =\left(\begin{array}{cccc}
    -1&-1&-1&-1\\
    \frac{a}{\piff}&\frac{b+w_0}{\pifb}&
    \frac{ba+w_0u_1}{(b+w_0)\pibf}&\frac{b}{\pibb}
  \end{array}\right).
\end{multline}
Solving this and using $\rho_{\pm\pm}(t)=\partial_tP_{\pm\pm}(t)$, we
eventually arrive at the probability density functions given in Eqs.\
(\ref{ffbbExample}-\ref{fbbfExample}).

Finally, we need expressions for the splitting probabilities, which we
derive using the adjoint equations \citep{vankampen}.  Let
$\pi_{j\pm}$, with $j=0,1$, be the probability that a system starting
in state $j$ will next produce a $\pm$ step.  Starting with
$\pi_{0-}$, we note that from state $0$, a backward step can be
accomplished either by the next transition being a backward step,
which has probability $(b+w_0)/(a+b+u_0+w_0)$, or by jumping to state
$1$ and from there eventually get a backward step. The last
possibility has total probability $\pi_{1-}u_0/(a+b+u_0+w_0)$, so that
\begin{equation}
  \pi_{0-}=\frac{b+w_0}{a+b+u_0+w_0}+\frac{u_0}{a+b+u_0+w_0}\pi_{1-}.
\end{equation}
Next, we multiply with $\sum_kw_{0k}=a+b+w_0+u_0$ and rearrange the
terms. Applying the same procedure for the other $\pi_{j\pm}$, we
finally get
\begin{equation}
  \matris{M}^T
  \left(\begin{array}{cc}\pi_{0-}&\pi_{0+}\\\pi_{1-}&\pi_{1+}\end{array}\right)
  =-\left(\begin{array}{cc}b+w_0&a\\0&u_1\end{array}\right).
\end{equation}
To get the pairwise splitting probabilities in Eqs.\
(\ref{ffbbExample}-\ref{N2Xgamma}), we first solve for the
$\pi_{j\pm}$, and then weight them according to the initial conditions
of \Eq{Aic}:
\begin{equation}
  \pi_{\pm+}=\sum_j\pi_{j+}q_j^{(\pm)}(0),\quad
    \pi_{\pm-}=\sum_j\pi_{j-}q_j^{(\pm)}(0).
\end{equation}
\end{appendix}
\begin{acknowledgments}
We are grateful to Richard Berry and Yoshiyuki Sowa for helpful
comments about the stepping data and flagellar motors in general, to
Michael E. Fisher, Denis Tsygankov and R. Dean Astumian for
stimulating discussions, and to Hong Qian for sharing Ref.\
\citep{wang06} with us prior to publication. Financial support from
the Royal Institute of Technology, the Wallenberg Foundation (M.L.)
and the Swedish Research Council, grant 2003-5001, (M.W.) is
gratefully acknowledged.
\end{acknowledgments}


\begin{thebibliography}{58}
\providecommand{\url}[1]{\texttt{#1}}
\providecommand{\urlprefix}{ }

\bibitem[Carter and Cross(2005)]{carter05}
Carter, N.~J., and R.~A. Cross, 2005.
\newblock Mechanics of the kinesin step.
\newblock \emph{Nature} 435:308--312.

\bibitem[Nishiyama et~al.(2002)Nishiyama, Higuchi, and Yanagida]{nishiyama02}
Nishiyama, M., H.~Higuchi, and T.~Yanagida, 2002.
\newblock Chemomechanical coupling of the forward and backward steps of single
  kinesin molecules.
\newblock \emph{Nature Cell Biol.} 4:790--797.

\bibitem[Taniguchi et~al.(2005)Taniguchi, Nishiyama, Ishii, and
  Yanagida]{taniguchi05}
Taniguchi, Y., M.~Nishiyama, Y.~Ishii, and T.~Yanagida, 2005.
\newblock Entropy rectifies the Brownian steps of kinesin.
\newblock \emph{Nat. Chem. Biol.} 1:342--347.

\bibitem[Guydosh and Block(2006)]{guydosh06}
Guydosh, N.~R., and S.~M. Block, 2006.
\newblock Backsteps induced by nucleotide analogs suggest the front head of
  kinesin is gated by strain.
\newblock \emph{Proc. Natl. Acad. Sci. U.S.A.} 103:8054--8059.

\bibitem[Uemura et~al.(2004)Uemura, Higuchi, Olivares, {De La Cruz}, and
  Ishiwata]{uemura04}
Uemura, S., H.~Higuchi, A.~O. Olivares, E.~M. {De La Cruz}, and S.~Ishiwata,
  2004.
\newblock Mechanochemical coupling of two substeps in a single myosin {V}
  motor.
\newblock \emph{Nat. Struct. Mol. Biol.} 11:877--883.

\bibitem[Rief et~al.(2000)Rief, Rock, Mehta, Mooseker, Cheney, and
  Spudich]{rief00}
Rief, M., R.~S. Rock, A.~D. Mehta, M.~S. Mooseker, R.~E. Cheney, and J.~A.
  Spudich, 2000.
\newblock Myosin-{V} stepping kinetics: A molecular model for processivity.
\newblock \emph{Proc. Natl. Acad. Sci.} 97:9482--9486.

\bibitem[Clemen et~al.(2005)Clemen, Vilfan, {Junshan Zhang}, B{\"a}rmann, and
  Rief]{clemen05}
Clemen, A. E.-M., M.~Vilfan, J.~J. {Junshan Zhang}, M.~B{\"a}rmann, and
  M.~Rief, 2005.
\newblock Force-Dependent Stepping Kinetics of Myosin-{V}.
\newblock \emph{Biophys. J.} 88:4401--4410.

\bibitem[Christof et~al.(2006)Christof, Gebhardt, Clemen, Jaud, , and
  Rief]{gebhardt06}
Christof, J., M.~Gebhardt, A.~E.-M. Clemen, J.~Jaud, , and M.~Rief, 2006.
\newblock {Myosin-V} is a mechanical ratchet.
\newblock \emph{Proc. Natl. Acad. Sci. U.S.A.} 103:8680--8685.

\bibitem[Toba et~al.(2006)Toba, Watanabe, Yamaguchi-Okimoto, Toyoshima, and
  Higuchi]{toba06}
Toba, S., T.~M. Watanabe, L.~Yamaguchi-Okimoto, Y.~Y. Toyoshima, and
  H.~Higuchi, 2006.
\newblock Overlapping hand-over-hand mechanism of single molecular motility of
  cytoplasmic dynein.
\newblock \emph{Proc. Natl. Acad. Sci. U.S.A.} 103:5741--5745.

\bibitem[Abbondanzieri et~al.(2005)Abbondanzieri, Greenleaf, Shaevitz, Landick,
  and Block]{abbondanzieri05}
Abbondanzieri, E.~A., W.~J. Greenleaf, J.~W. Shaevitz, R.~Landick, and S.~M.
  Block, 2005.
\newblock Direct observation of base-pair stepping by {RNA} polymerase.
\newblock \emph{Nature} 438:460--465.

\bibitem[Diez et~al.(2004)Diez, Zimmermann, Börsch, König, Schweinberger,
  Steigmiller, Reuter, Felekyan, Kudryavtsev, Seidel, and Gräber]{diez04}
Diez, M., B.~Zimmermann, M.~Börsch, M.~König, E.~Schweinberger, S.~Steigmiller,
  R.~Reuter, S.~Felekyan, V.~Kudryavtsev, C.~A.~M. Seidel, and P.~Gräber, 2004.
\newblock Proton-powered subunit rotation in single membrane-bound
  F$_0$F$_1$-{ATP} synthase.
\newblock \emph{Nat. Struct. Mol. Biol.} 11:135--141.

\bibitem[Ueno et~al.(2005)Ueno, Suzuki, {Kinosita Jr.}, , and Yoshida]{ueno05}
Ueno, H., T.~Suzuki, K.~{Kinosita Jr.}, , and M.~Yoshida, 2005.
\newblock {ATP}-driven stepwise rotation of F$_0$F$_1$-{ATP} synthase.
\newblock \emph{Proc. Natl. Acad. Sci. U.S.A.} 102:1333--1338.

\bibitem[Yasuda et~al.(1998)Yasuda, Noji, {Kinosita Jr.}, and
  Yoshida]{yasuda98}
Yasuda, R., H.~Noji, K.~{Kinosita Jr.}, and M.~Yoshida, 1998.
\newblock F$_1$-ATPase Is a Highly Efficient Molecular Motor that Rotates with
  Discrete 120° Steps.
\newblock \emph{Cell} 93:1117--1124.

\bibitem[Yasuda et~al.(2001)Yasuda, Noji, Yoshida, {Kinosita Jr.}, and
  Itoh]{yasuda01}
Yasuda, R., H.~Noji, M.~Yoshida, K.~{Kinosita Jr.}, and H.~Itoh, 2001.
\newblock Resolution of distinct rotational substeps by submillisecond kinetic
  analysis of F1-ATPase.
\newblock \emph{Nature} 410:898--904.

\bibitem[Shimabukuro et~al.(2003)Shimabukuro, Yasuda, Muneyuki, Hara, {Kinosita
  Jr.}, and Yoshida]{shimabukuro03}
Shimabukuro, K., R.~Yasuda, E.~Muneyuki, K.~Y. Hara, K.~{Kinosita Jr.}, and
  M.~Yoshida, 2003.
\newblock Catalysis and rotation of F$_1$ motor: Cleavage of ATP at the
  catalytic site occurs in 1 ms before $40^\circ$ substep rotation.
\newblock \emph{Proc. Natl. Acad. Sci. U.S.A.} 100:14731--14736.

\bibitem[Shimabukuro et~al.(2005)Shimabukuro, Muneyuki, and
  Yoshida]{shimabukuro05}
Shimabukuro, K., E.~Muneyuki, and M.~Yoshida, 2005.
\newblock An Alternative Reaction Pathway of F1-ATPase Suggested by Rotation
  without $80^o/40^o$ Substeps of a Sluggish Mutant at Low {ATP}.
\newblock \emph{Biophys. J.} 90:1028--1032.

\bibitem[Nishizaka et~al.(2004)Nishizaka, Oiwa, Noji, Kimura, Muneyuki,
  Yoshida, and {Kinosita Jr}]{nishizaka04}
Nishizaka, T., K.~Oiwa, H.~Noji, S.~Kimura, E.~Muneyuki, M.~Yoshida, and
  K.~{Kinosita Jr}, 2004.
\newblock Chemomechanical coupling in F$_1$-ATPase revealed by simultaneous
  observation of nucleotide kinetics and rotation.
\newblock \emph{Nat. Struct. Mol. Biol.} 11:142--148.

\bibitem[Sowa et~al.(2005)Sowa, Rowe, Leake, Yakushi, Homma, Ishijima, and
  Berry]{sowa05}
Sowa, Y., A.~D. Rowe, M.~C. Leake, T.~Yakushi, M.~Homma, A.~Ishijima, and R.~M.
  Berry, 2005.
\newblock Direct observation of steps in rotation of the bacterial flagellar
  motor.
\newblock \emph{Nature} 437:916--919.

\bibitem[{van Kampen}(1992)]{vankampen}
{van Kampen}, N.~G., 1992.
\newblock Stochastic Processes in Physics and Chemistry.
\newblock Elsevier Science Publishers {B.V.}, Amsterdam, The Netherlands, 2nd
  edition.

\bibitem[Tsygankov et~al.(2007)Tsygankov, Lind\'en, and Fisher]{tsygankov06a}
Tsygankov, D., M.~Lind\'en, and M.~E. Fisher, 2007.
\newblock Hidden substeps in the catalytic cycle of molecular motors:
  conditional step probabilities and dwell times.
\newblock \emph{Phys. Rev. E} 75:021909.

\bibitem[Qian(1997)]{qian97}
Qian, H., 1997.
\newblock A simple theory of motor protein kinetics and energetics.
\newblock \emph{Biophys. Chem.} 67:263 -- 267.

\bibitem[Kolomeisky and Fisher(2003)]{kolomeisky03}
Kolomeisky, A.~B., and M.~E. Fisher, 2003.
\newblock A Simple Kinetic Model Describes the Processivity of Myosin-{V}.
\newblock \emph{Biophys. J.} 84:1642--1650.

\bibitem[Kolomeisky et~al.(2005)Kolomeisky, Stukalin, and Popov]{kolomeisky05}
Kolomeisky, A.~B., E.~B. Stukalin, and A.~A. Popov, 2005.
\newblock Understanding mechanochemical coupling in kinesins using
  first-passage-time processes.
\newblock \emph{Phys. Rev. E} 71:031902.

\bibitem[Qian and {Sunney Xie}(2006)]{qian06}
Qian, H., and X.~{Sunney Xie}, 2006.
\newblock Generalized Haldane Equation and Fluctuation Theorem in the Steady
  State Cycle Kinteics of Single Enzymes.
\newblock \emph{Phys. Rev. E} 74:010902(R).

\bibitem[Wang and Qian(2007)]{wang06}
Wang, H., and H.~Qian, 2007.
\newblock On Detailed Balance and Reversibility of Semi-Markov Processes and
  Single-molecule Enzyme Kinetics.
\newblock \emph{J. Math. Phys.} 48:013303.

\bibitem[Shaevitz et~al.(2005)Shaevitz, Block, and Schnitzer]{shaevitz05}
Shaevitz, J.~W., S.~M. Block, and M.~J. Schnitzer, 2005.
\newblock Statistical Kinetics of Macromolecular Dynamics.
\newblock \emph{Biophys. J.} 89:2277--2285.

\bibitem[Santos et~al.(2005)Santos, Franosch, Parmeggiani, and Frey]{santos05}
Santos, J.~E., T.~Franosch, A.~Parmeggiani, and E.~Frey, 2005.
\newblock Renewal processes and fluctuation analysis of molecular motor
  stepping.
\newblock \emph{Phys. Biol.} 2:207--222.

\bibitem[Svoboda et~al.(1994)Svoboda, Mitra, and Block]{svoboda94}
Svoboda, K., P.~P. Mitra, and S.~M. Block, 1994.
\newblock Fluctuation analysis of Motor Protein Movement and Single Enzyme
  Kinetics.
\newblock \emph{Proc. Natl. Acad. Sci.} 91:11782--11786.

\bibitem[Howard(2001)]{howard}
Howard, J., 2001.
\newblock Mechanics of Motor Proteins and the Cytoskeleton.
\newblock Sinauer Ass. Inc., Sunderland M.A., U.S.A.

\bibitem[Bustamante et~al.(2001)Bustamante, Keller, and Oster]{bustamante01}
Bustamante, C., D.~Keller, and G.~Oster, 2001.
\newblock The Physics of Molecular Motors.
\newblock \emph{Acc. Chem. Res.} 34:412--420.

\bibitem[Itoh et~al.(2004)Itoh, Takahashi, Adachi, Noji, Yasuda, Yoshida, and
  {Kinosita, Jr}]{itoh04}
Itoh, H., A.~Takahashi, K.~Adachi, H.~Noji, R.~Yasuda, M.~Yoshida, and
  K.~{Kinosita, Jr}, 2004.
\newblock Mechanically driven {ATP} synthesis by F$_1$-ATPase.
\newblock \emph{Nature} 427:465--468.

\bibitem[Berg(2000)]{berg00}
Berg, H.~C., 2000.
\newblock Constraints on Models for the Flagellar Rotary Motor.
\newblock \emph{Phil. Trans. R. Soc. Lond. B.} 355:491--501.

\bibitem[Milescu et~al.(2006{\natexlab{a}})Milescu, Yildiz, Selvin, and
  Sachs]{milescu06}
Milescu, L.~S., A.~Yildiz, P.~R. Selvin, and F.~Sachs, 2006.
\newblock Maximum Likelihood Estimation of Molecular Motor Kinetics from
  Staircase Dwell-Time Sequences.
\newblock \emph{Biophys. J.} 91:1156--1168.

\bibitem[Milescu et~al.(2006{\natexlab{b}})Milescu, Yildiz, Selvin, and
  Sachs]{milescu06b}
Milescu, L.~S., A.~Yildiz, P.~R. Selvin, and F.~Sachs, 2006.
\newblock Extracting Dwell Time Sequences from Processive Molecular Motor Data.
\newblock \emph{Biophys. J.} 91:3135--3150.

\bibitem[McKinney et~al.(2006)McKinney, Joo, and Ha]{mckinney06}
McKinney, S.~A., C.~Joo, and T.~Ha, 2006.
\newblock Analysis of Single-Molecule FRET Trajectories Using Hidden Markov
  Modeling.
\newblock \emph{Biophys. J.} 91:1941--1951.

\bibitem[Smith et~al.(2001)Smith, Steffen, Simmons, and Sleep]{smith01}
Smith, D.~A., W.~Steffen, R.~M. Simmons, and J.~Sleep, 2001.
\newblock Hidden-Markov Methods for the Analysis of Single-Molecule Actomyosin
  Displacement Data: The Variance-Hidden-Markov Method.
\newblock \emph{Biophys. J.} 81:2795--2816.

\bibitem[Hill(1989)]{hill89}
Hill, T.~L., 1989.
\newblock Free Energy Transduction and Biochemical Cycle Kinetics.
\newblock Springer-Verlag, New York, U.S.A.

\bibitem[Fisher and Kolomeisky(2001)]{fisher01}
Fisher, M.~E., and A.~B. Kolomeisky, 2001.
\newblock Simple mechanochemistry describes the dynamics of kinesin molecules.
\newblock \emph{Proc. Natl. Acad. Sci. U.S.A.} 98:7748--7753.

\bibitem[Fisher and Kim(2005)]{fisher05}
Fisher, M.~E., and Y.~C. Kim, 2005.
\newblock Kinesin crouches to sprint but resist pushing.
\newblock \emph{Proc. Natl. Acad. Sci.} 102:16209--16214.

\bibitem[Gao et~al.(2005)Gao, Yang, and Karplus]{gao05}
Gao, Y., W.~Yang, and M.~Karplus, 2005.
\newblock A Structure-Based Model for the Synthesis and Hydrolysis of ATP by
  F1-ATPasein this model.
\newblock \emph{Cell} 123:195--205.

\bibitem[Fisher and Kolomeisky(1999{\natexlab{a}})]{fisher99b}
Fisher, M.~E., and A.~B. Kolomeisky, 1999.
\newblock The force exerted by a molecular motor.
\newblock \emph{Proc. Natl. Acad. Sci. U.S.A.} 96:6597--6602.

\bibitem[Wang et~al.(2003)Wang, Peskin, and Elston]{wang03}
Wang, H., C.~S. Peskin, and T.~C. Elston, 2003.
\newblock A Robust Numerical Algorithm for Studying Biomolecular Transport
  Processes.
\newblock \emph{J. Theor. Biol.} 221:491--511.

\bibitem[Bier et~al.(1999)Bier, Derényi, Kostur, and Astumian]{bier99}
Bier, M., I.~Derényi, M.~Kostur, and R.~D. Astumian, 1999.
\newblock Intrawell relaxation of overdamped Brownian particles.
\newblock \emph{Phys. Rev. E} 59:6422--6432.

\bibitem[Kolomeisky(2001)]{kolomeisky01}
Kolomeisky, A.~B., 2001.
\newblock Exact results for parallel-chain kinetic models of biological
  transport.
\newblock \emph{J. Chem. Phys.} 115:7253--7259.

\bibitem[Baker et~al.(2004)Baker, Krementsova, Kennedy, Armstrong, Trybus, and
  Warshaw]{baker04}
Baker, J.~E., E.~B. Krementsova, G.~G. Kennedy, A.~Armstrong, K.~M. Trybus, and
  D.~M. Warshaw, 2004.
\newblock Myosin {V} processivity: Multiple kinetic pathways for head-to-head
  coordination.
\newblock \emph{Proc. Natl. Acad. Sci. U.S.A.} 101:5542--5546.

\bibitem[Vilfan(2005)]{vilfan05}
Vilfan, A., 2005.
\newblock Elastic Lever-Arm Model for Myosin V.
\newblock \emph{Biophys. J.} 88:3792--3805.

\bibitem[Lan and Sun(2005)]{lan05}
Lan, G., and S.~X. Sun, 2005.
\newblock Dynamics of Myosin-V Processivity.
\newblock \emph{Biophys. J.} 88:999--1008.

\bibitem[Xing et~al.(2005)Xing, Liao, and Oster]{xing05}
Xing, J., J.-C. Liao, and G.~Oster, 2005.
\newblock Making ATP.
\newblock \emph{Proc. Natl. Acad. Sci. U.S.A.} 102:16539--16546.

\bibitem[Reimann(2002)]{reimann02}
Reimann, P., 2002.
\newblock Brownian motors: noisy transport far from equilibrium.
\newblock \emph{Phys. Rep.} 361:57--265.

\bibitem[J{\"u}licher et~al.(1997)J{\"u}licher, Ajdari, and Prost]{juelicher97}
J{\"u}licher, F., A.~Ajdari, and J.~Prost, 1997.
\newblock Modeling molecular motors.
\newblock \emph{Rev. Mod. Phys.} 69:1269--1282.

\bibitem[Kolomeisky and Fisher(2000)]{kolomeisky00}
Kolomeisky, A.~B., and M.~E. Fisher, 2000.
\newblock Periodic sequential kinetic models with jumping, branching and
  deaths.
\newblock \emph{Physica A} 279:1--20.

\bibitem[Berg(2003)]{berg03}
Berg, H.~C., 2003.
\newblock The rotary motor of bacterial flagella.
\newblock \emph{Annu. Rev. Biochem.} 72:19--54.

\bibitem[Chung and Kennedy(1991)]{chung91}
Chung, S.~H., and R.~A. Kennedy, 1991.
\newblock Forward-Backward non-linear filtering technique for extracting small
  biological signals from noise.
\newblock \emph{J. Neurosci. Methods} 40:71--86.

\bibitem[Reid et~al.(2006)Reid, Leake, Chandler, Lo, Armitage, and
  Berry]{reid06}
Reid, S.~W., M.~C. Leake, J.~H. Chandler, C.-J. Lo, J.~P. Armitage, and R.~M.
  Berry, 2006.
\newblock The maximum number of torque-generating units in the flagellar motor
  of Escherichia coli is at least 11.
\newblock \emph{Proc. Natl. Acad. Sci. U.S.A.} 103:8066--8071.

\bibitem[Press et~al.(1992)Press, Teukolsky, Vetterling, and Flannery]{numres}
Press, W.~H., S.~A. Teukolsky, W.~T. Vetterling, and B.~P. Flannery, 1992.
\newblock Numerical recipes in C.
\newblock Cambridge University Press, 2 edition.

\bibitem[Barlow(1989)]{barlow}
Barlow, R.~J., 1989.
\newblock Statistics.
\newblock Wiley, Chichester, England.

\bibitem[Koza(2002)]{koza02}
Koza, Z., 2002.
\newblock Maximal force exerted by a molecular motor.
\newblock \emph{Phys. Rev. E} 65:031905.

\bibitem[Fisher and Kolomeisky(1999{\natexlab{b}})]{fisher99a}
Fisher, M.~E., and A.~B. Kolomeisky, 1999.
\newblock Molecular motors and the forces they exert.
\newblock \emph{Physica A} 274:241--266.

\end{thebibliography}

\end{document}